\documentclass[universe,review,accept,pdftex,moreauthors]{Definitions/mdpi} 

\firstpage{1} 
\pubvolume{1}
\issuenum{1}
\articlenumber{0}
\pubyear{2024}
\copyrightyear{2024}
\externaleditor{Academic Editor:}
\datereceived{30 June 2024} 
\daterevised{30 July 2024} 
\dateaccepted{1 August 2024} 
\datepublished{ } 
\hreflink{https://doi.org/} 
\usepackage{comment}
\usepackage{caption}
\usepackage{subcaption}
\usepackage{longtable}

\Title{Multi-Messenger Connection in High-Energy \mbox{Neutrino Astronomy}}

\TitleCitation{Multi-Messenger Connection in High-Energy Neutrino Astronomy}

\Author{Ankur Sharma $^{1,2}$\orcidA{}}

\AuthorNames{Ankur Sharma}

\AuthorCitation{Sharma, A.}

\address{%
$^{1}$ \quad UFR Physics, Université Paris Cité, Bâtiment Condorcet, 10 rue Alice Domon et Léonie Duquet, 75013 Paris, France   \\
$^{2}$ \quad Laboratoire Astroparticule et Cosmologie (APC), Bâtiment Condorcet, 10 rue Alice Domon et Léonie Duquet, 75013 Paris, France; \\
\hspace{11pt} \href{mailto:sharma@apc.in2p3.fr}{sharma@apc.in2p3.fr}
} 

\abstract{Low fluxes of astrophysical neutrinos at TeV energies, and the overwhelming background of atmospheric neutrinos below that, render the current paradigm of neutrino astronomy a severely statistics-limited one.  While many hints have emerged, all the evidence gathered by IceCube and ANTARES, over the course of almost a decade and a half of operation, has fallen short of providing any conclusive answer to the puzzle of the origin of high-energy cosmic rays and neutrinos. The advancement of the field is thus closely associated with not only the neutrino observatories coming online in the next few years, but also on the coordinated efforts of the EM, GW and cosmic ray communities to develop dedicated channels and infrastructure that allow for the swift and comprehensive multi-messenger follow-up of relevant events detected in any of these sectors. This paper highlights the strides that have been already taken in that direction and the fruits that they have borne, as well as the challenges that lie ahead.}


\DeclareUnicodeCharacter{2212}{-}
\begin{document}


\section{Introduction}

Multi-messenger astronomy represents a transformative approach to understanding the universe, where information about astronomical events is gathered through various types of ``messengers'' such as photons (electromagnetic radiation), neutrinos, cosmic rays and~gravitational waves. The~field, while still in its infancy, has seen a rapid expansion, leveraging the unique insights provided by each messenger to build a more complete picture of the universe's most energetic phenomena. 

Multi-messenger observations were already being put to use as early as 1987 to confirm the first extra-galactic source seen in neutrinos, the~supernova SN1987A~\cite{1987PhRvL..58.1490H, 1987PZETF..45..461A}. And~possibly even earlier, if~we account for the decades of observations of our Sun that led to the prediction of solar neutrino flux~\cite{sun_nu} and the subsequent hunt for neutrinos~\cite{Kuzmin:1965zza, Davis:1968cp, Cleveland:1998nv}. However, the~recent resurgence of the field can be traced back to the discovery of the first gravitational wave event connected to the merger of two stellar-mass binary BHs in 2015~\cite{gw_first}, which was followed up throughout the electromagnetic spectrum and confirmed the existence of BHs with $M_{\odot} > 25$. A~mere two years later in 2017, the~IceCube neutrino observatory detected a multi-TeV neutrino from the direction of a blazar that was found to be flaring in gamma-rays by Fermi-LAT~\cite{txs_ic}, {while a binary neutron star merger was also identified through the GRB counterpart to a LIGO alert in the same year~\cite{ligo_2017}}. This was a significance boost for multi-messenger efforts, since it demonstrated the effectiveness of making these observations in real time to enhance the probability of astrophysical discoveries. 

The central promise of multi-messenger astronomy with neutrinos lies in it being able to narrow down the search for hadronic accelerators by searching in a small, almost background-free, window of space and/or time. With~proton astronomy feasible only at the highest energies, and~the large localization in space provided by the currently operational gravitational-wave observatories, EM radiation becomes the messenger of choice to complement or direct high-energy neutrino searches. The~next few sections detail the theoretical motivations and current status of this emerging field and outline the challenges lying along the way. We conclude by looking at the future prospects for this broad field and by shedding light on the possibilities provided by the planned and upcoming EM and neutrino~observatories.  


\section{The~Basics}

\begin{figure}[!hbt] 

\includegraphics[width=12cm]{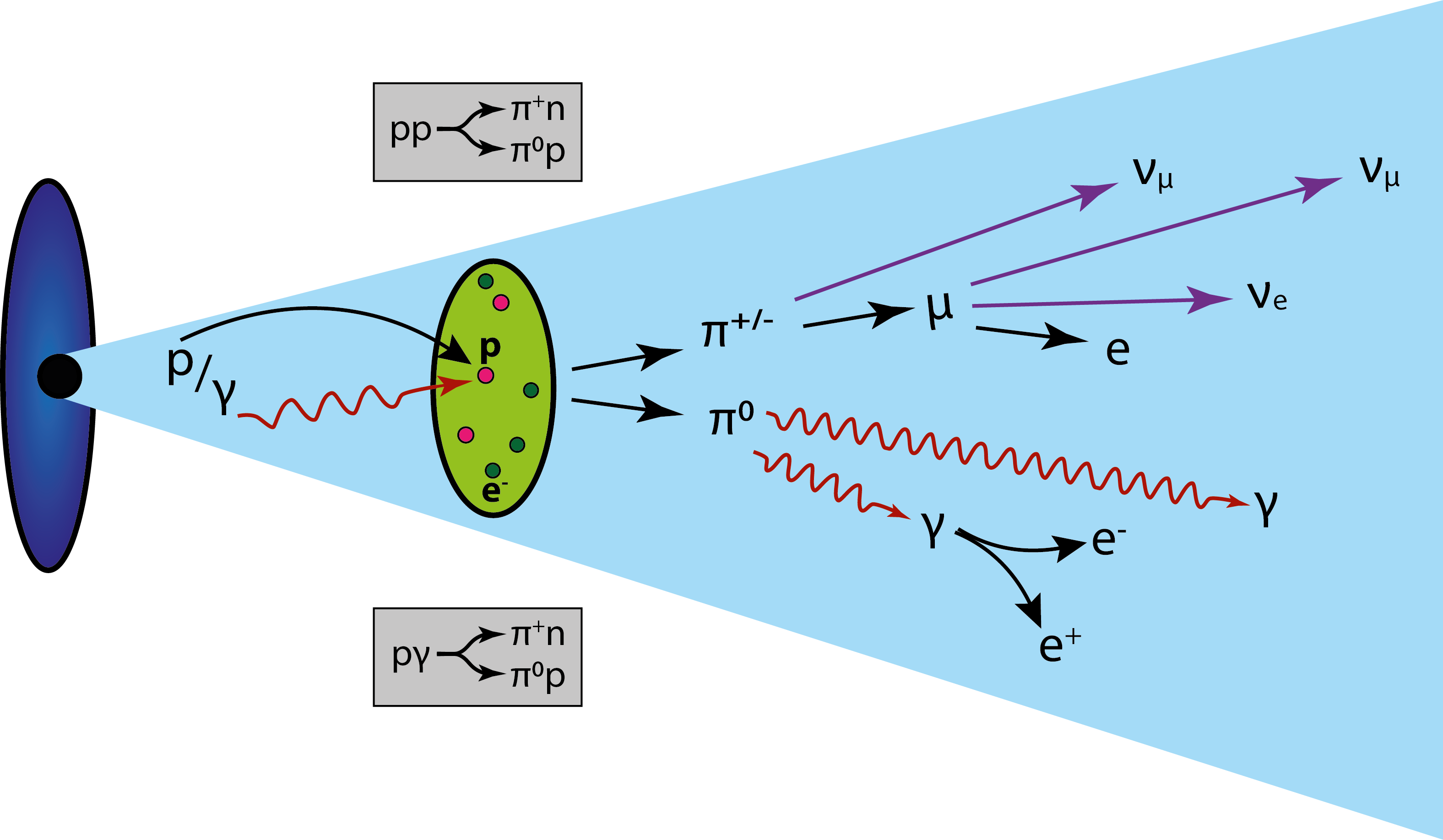}
\caption{The high-energy neutrino production channels present in astrophysical environments, shown here for a jetted galaxy. Protons are co-accelerated with electrons (shown here inside a blob) and can interact with ambient photons or cold gas. Both \textit{pp} and $p\gamma$ interactions lead to the production of charged and neutral pions, which further decay to produce $\gamma$-rays and neutrinos. Neutrinos escape the environment mostly unhindered, while the $\gamma$-rays can be absorbed internally and cascade down in energy through the production of secondary~pairs. }
\label{fig:nu_prod}
\end{figure}

In astrophysical environments, accretion processes release gravitational energy, which is stored in the magnetic field arising from a central compact object. These field lines, in~turn, {can help accelerate cosmic ray nuclei and protons to ultra-relativistic energies through diffusive shock acceleration or magnetic reconnection processes~\cite{mag_reconnect, fermi_shock}}. Shock fronts of supernovae also provide favorable conditions for diffusive shock acceleration. These~accelerated cosmic rays can interact with either the photons produced by electrons co-accelerated in this strong magnetic field (electron synchrotron) or~with external radiation from the surrounding region, leading to the production of high-energy neutrinos through \textit{photohadronic} 
($p\gamma$) interactions (see Equation (\ref{eqn:eqn_pg})). If~the accelerated protons come into contact with cold gas from the inter-stellar medium, $pp$ collisions (\textit{an astrophysical beam dump}) result, again producing neutrinos as an end product. While $p\gamma$ interactions produce charged and neutral pions {in a ratio of 1/3 to 2/3~\cite{pi_ratio}}, an~equal number of charged and neutral pions are expected from the beam dump process. The~$\pi^0 \rightarrow 2 \gamma$, while the charged pions decay in the manner described in Equation (\ref{eqn:piplus_decay}) (see also Figure~\ref{fig:nu_prod}). In~their final state, $\sim$10\% of the energy of the parent proton/nuclei on average is carried away by each $\gamma$, while each neutrino shares $\sim$5\% of its parent nucleon's energy. The~spectrum of gamma-rays at the source is thus dictated by the primary protons and shares the same spectral index as the~latter.   

\vspace{-12pt}    

\begin{linenomath}
\begin{equation} \label{eqn:eqn_pg}
\begin{split}
p + \gamma \rightarrow \Delta^+ \rightarrow p + \pi^0 \\
                                \rightarrow n + \pi^+ 
\end{split}
\end{equation}
\end{linenomath}
\vspace{-12pt} 
\begin{linenomath}
\begin{equation} \label{eqn:piplus_decay}
\begin{split}
 \pi^+ \rightarrow \mu^+ + \nu_{\mu} \\
 \mu^+ \rightarrow e^+ + \overline{\nu}_{\mu} + \nu_e
 \end{split}
\end{equation}
\end{linenomath}

While these generation processes dictate comparable fluxes at the source for gamma-rays and neutrinos, the~evidence of hadronically produced gamma-rays from an astrophysical source is, so far, scarce~\cite{hadronic_gamma1, hadronic_gamma2, hadronic_gamma3, hadronic_gamma4, hadronic_gamma5, hadronic_gamma6}. Efficient neutrino production proceeds in the presence of a dense target field. However, the~same target radiation can be responsible for the absorption of gamma-rays inside the source, leading to the production of secondaries, which cascade down their energy until the source becomes transparent to them. This redistributed energy from the TeV--PeV range appears in the X-ray to GeV band~\cite{em_cascades1, em_cascades2}, where observational efforts can be focused to constrain the expected neutrino flux from a source. {In recent years, this multi-messenger connection has been increasingly exploited. If~a source shows increased activity at any wavelength, or~a neutrino event of high purity is detected from its direction, it is observed simultaneously in multiple wavelength bands.} Correlations in non-thermal emissions, from~soft X-rays to VHE $\gamma$-rays, provide hints on the particle populations involved in the jet and accretion region interactions (or shock front interactions in the case of SNRs), while radio variability can act as a tracer of the central compact object's activity, 
thus indicating the location of the emission region relative to the galaxy~\cite{ic_radio_nu}.   

In addition, the~EBL further depletes the gamma-ray flux at energies of $\mathcal{O}$(100 GeV) and above depending on the redshift~\cite{Franceschini_2008, Finke_2010, dominguez_thesis}. This steep cutoff further complicates the simultaneous detection of TeV gamma-rays produced alongside the TeV neutrinos observed by Earth-based neutrino detectors~\cite{Singh_2020}.  


\section{Real-time Alert~Programs} \label{sec:realtime}

The statistics-limited sample of astrophysical neutrinos in the TeV regime and above, coupled with the angular resolution of existing neutrino telescopes, implies that individual events cannot reliably establish the origin of their sources. The~typical angular uncertainty above 100 TeV is <1$^{\circ}$ for state-of-the-art detectors and~is constantly improving with better instrumentation and reconstruction algorithms {\cite{km3net_loi, ic_angular_reco}}. However, multiple or zero point-sources within the containment region is a common occurrence. 

An optimal strategy to alleviate the problem of identifying the possible source of origin of a high-energy neutrino event is to employ the EM observatories to search for potential counterparts to the event. This is achieved by broadcasting alerts for signal-like events to the astronomy community in real time (latency < 1 min) with the help of brokers. Currently, only IceCube has an active alert broadcast program in place~\cite{ic_realtime}, although~KM3NeT~\cite{km3net_loi} is expected to join it soon~\cite{Celli:2023tG}. Baikal-GVD~\cite{baikal} has an alert follow-up program to look for coincidences in real time in response to external triggers, like those from IceCube, with their own alert pipeline in its planning phase~\cite{Dik_2023, allakhverdyan2021multimessenger}. The~predecessor to KM3NeT, the~ANTARES observatory~\cite{antares}, also had an alert follow-up program in place until its decommission in 2022~\cite{albert2024results}, with~several external GRB alerts followed up. 

\begin{figure}[H] 

\includegraphics[width=12cm]{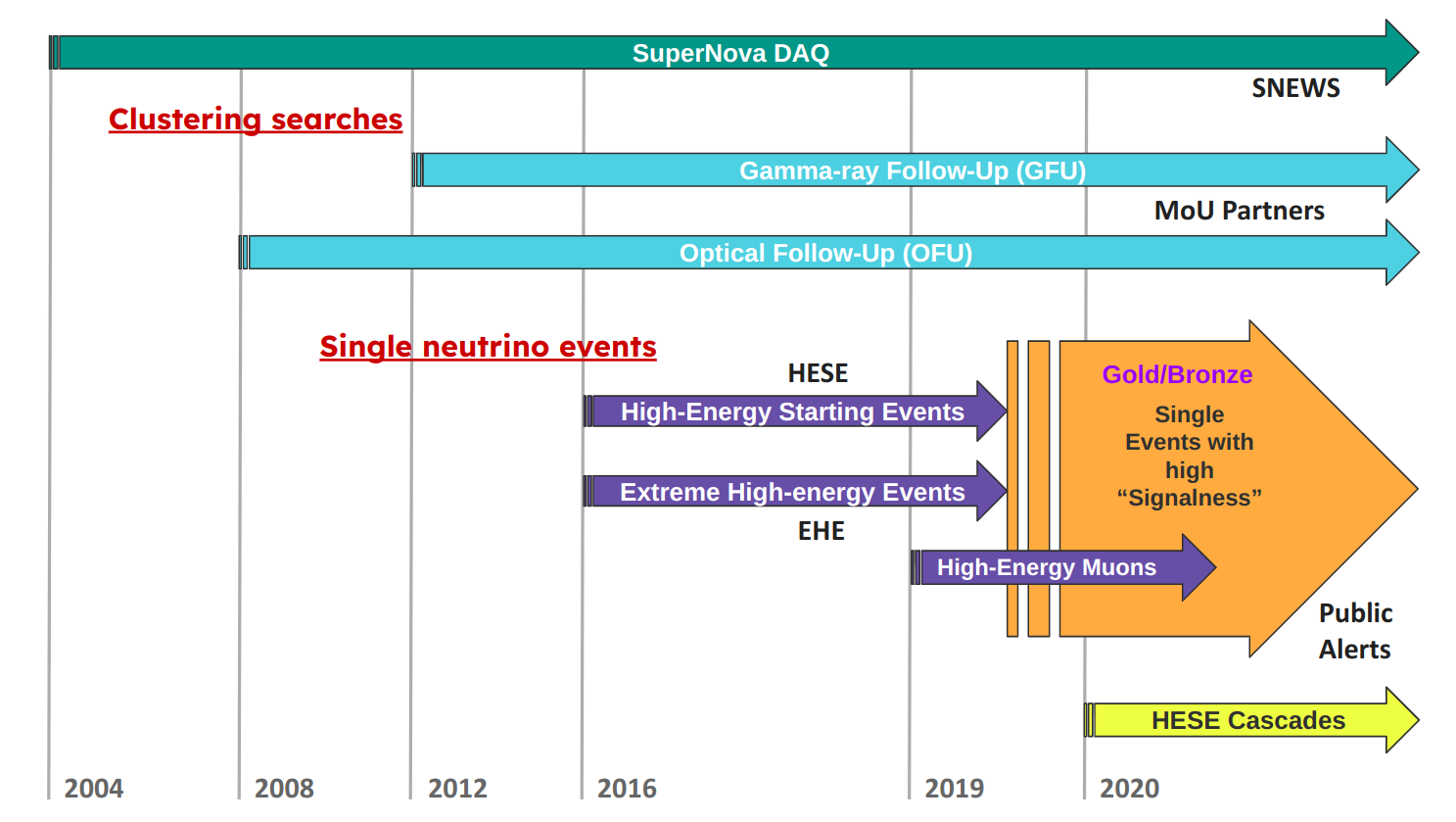}
\caption{The public and private alert streams in IceCube for broadcasting neutrino alerts. SNEWS has existed in some form since around 2004--2005 for coordinating early SN warning follow-ups. The GFU and OFU alerts are sent out privately to the MoU-partnering experiments, while single track-like alerts are publicly announced. Alerts for starting HESE cascade events have also been made available in the public domain since~2020.}
\label{fig:ic_alert_streams}
\end{figure}

IceCube's real-time alert program was established in 2016, with~its HESE and EHE streams for individual track alerts brokered to the public through GCN~\cite{gcn} and AMON~\cite{amon}. However, an~MoU-based stream for private alerts to gamma-ray observatories, for~accumulated significance over a time window from pre-defined sky directions (or an unmonitored part of the sky), called GFU, was already in place to monitor or detect GeV-TeV bright sources~\cite{ic_realtime}. The~observatories currently under MoU to receive GFU alerts include HAWC~\cite{hawc}, MAGIC~\cite{magic}, VERITAS~\cite{veritas}, HESS~\cite{hess} and {Fermi} \cite{fermi_lat}. Additionally, the~OFU alerts were broadcast to the participating optical and X-ray observatories (SWIFT~\cite{swift}, MASTER network~\cite{master}, ZTF~\cite{ztf}, INTEGRAL~\cite{integral}, etc.) with the same event selection: a~multiplet of events within 100 s from an area of a radius $\sim$3.5$^{\circ}$ in the sky (Figure \ref{fig:ic_alert_streams}). 

Within less than 2 years of the operation of the IceCube public alert pipeline, an~EHE alert of $E_{\nu}$$\sim$290 TeV and an error of $\sim$$0.6^{\circ}$ was successfully followed up to reveal evidence of a neutrino-emitting blazar~\cite{txs_ic}, which prompted several investigations into the counterparts to these alert events. Since then, there have been further correlations between IceCube alerts and extra-galactic sources including TDEs~\cite{alerts_corr1, alerts_corr2, alerts_corr3, alerts_corr4, alerts_corr5}, although~none have resulted in a conclusive identification of a high-energy neutrino source. 

The streams for track alerts have been refined from their previous classification based on the starting events in the detector or through-going events above $\sim$200 TeV, and~are now labeled \textit{Gold} or \textit{Bronze} based on the \textit{signalness} of the events~\cite{ic_realtime}. An~average of $\sim$10 high-purity ($P_{astro} > 50\%$) alerts per year come out of these streams, which are used to train observatories, from the radio band (VLA~\cite{vla}, LOFAR~\cite{lofar}, etc.), to~the optical (ZTF, ASAS-SN network~\cite{assasn}, MASTER, etc.), to~X-rays (XMM-Newton~\cite{xmm}, NuSTAR~\cite{nustar, o2007optics}) and to gamma-rays (HESS, MAGIC, VERITAS, etc.), in~the direction of the event. IceCube has recently published a catalog of its alerts, which also includes the archival events detected before the establishment of the program that fit under the criteria of an individual track-like alert~\cite{icecat}. A~new stream for HESE cascade alerts was added recently~\cite{cascade_alerts}, and~the SNEWS 2.0 pipeline will soon upgrade the early supernova alert stream SNEWS~\cite{SNEWS}, to~be followed-up by neutrino, GW and optical observatories~\cite{SNEWS2}. 

In addition to EM follow-up by the community, each neutrino sub-threshold alert created by IceCube is followed up internally with the FRA, searching in windows of $\pm$500 s and $\pm$1~day around the neutrino event in the archival data. The~FRA is also triggered in response to external EM or GW alerts for transients like GRBs, TDEs, compact binary mergers, etc.~\cite{ic_fra}. 

EM observatories also issue alerts for transient events to the neutrino and GW communities. Neutrino follow-ups of EM alerts have the advantage of a significantly reduced background of atmospheric neutrinos due to the limited region of the sky and the small time window searched, increasing the probability of detecting an excess. GW alerts are also followed up by neutrino observatories, with Baikal-GVD and KM3NeT being involved in this exercise in addition to IceCube. No significant coincidence has been found so far by any of these neutrino telescopes.   

The exchange of information within the astronomical community is facilitated by brokers like GCN and AMON, through automated notices and human-readable circulars, which receive an alert distributed by the connected observatories, process the relevant information for brevity, and~make it available in real time to the interested parties for follow-up. In~view of the swiftly increasing volume of alerts expected from the upcoming facilities like Vera Rubin~\cite{lsst}, IceCube-Gen2~\cite{ic_gen2}, etc., and~the diversity of the stakeholders involved, the~role of brokers and common data-formats in alerts from observatories of a similar kind become crucial. Astro-COLIBRI~\cite{astro_colibri} has become a very comprehensive broker for alerts of all kinds in the last couple of years and it is still expanding its capabilities. GCN will be replaced by TACH~\cite{tach} in the next few years, while LSST and others will benefit from dedicated advanced alert brokers, like SciMMA~\cite{scimma} for the multi-messenger community. The~role of AI has also become important in these new platforms, which use LLMs to parse the text in circulars for parameters of interest.

\section{Experimental~Hints}

\begin{figure}[H] 

\includegraphics[width=12cm]{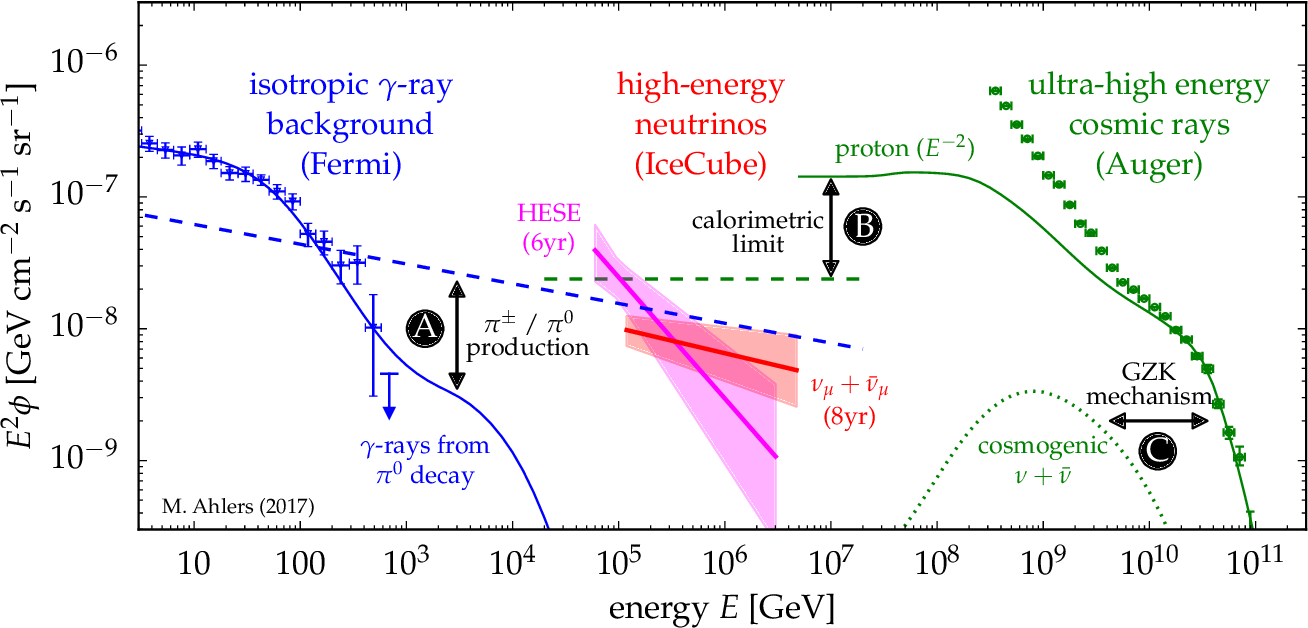}
\caption{Energy density of the common messengers. Neutrino flux is taken from the eight-year upgoing track analysis (red fit) and the six-year HESE analysis ({magenta points}) of IceCube. The~shaded region represents the $1 \sigma$ uncertainty. In~blue is the flux of unresolved extragalactic $\gamma$-ray sources~\cite{ahlers100} and in green is the flux of UHECRs from~\cite{ahlers101}. The~cosmogenic neutrino flux is also indicated with green dashed lines. For~more details, see the original paper~\cite{Ahlers_2018}.}
\label{fig:common_energy_density}
\end{figure}

The evidence for a diffuse TeV astrophysical neutrino flux, measured first in 2013 by IceCube, has grown to $\ge 6 \sigma$ level~\cite{abbasi2024characterization}. Pierre Auger, operational since early 2000s, has detected UHECRs reaching energies of several EeVs~\cite{Abreu_2022} and characterized them to be largely of extra-galactic origin~\cite{cr_anisotropy}. Meanwhile, the Fermi satellite, launched in 2008, has constantly monitored the $\gamma$-ray sky in the MeV$-$GeV band. An~important insight that jumps out upon putting together the data from these three different messengers is that the energy density in each of these particles, in~their respective energy ranges, is similar~\cite{Ahlers_2018}. This hints at a common origin and potentially a common production mechanism for the three (Figure \ref{fig:common_energy_density}). Calculations with simplistic assumptions hint at an energy injection density of $\sim$$10^{44}$--$10^{45}$ erg/Mpc$^3$/yr for neutrinos (between 0.1 and 1 PeV), UHECRs (>$10^{18}$ eV) and $\gamma$-rays (of 30 MeV--1 TeV) \cite{murase_41}.       

The isotropic gamma-ray background measured by Fermi-LAT can be used to constrain the sources of high-energy neutrinos. If~\textit{pp} collisions are the dominant channel for neutrino production in the universe, and~the sources are transparent to $\gamma$-rays, the~observed neutrino spectral index above 100 TeV cannot be > 2.1--2.2~\cite{Murase_44}. However, IceCube cascade data favors a softer index, and below 100 TeV, presents difficulties in reconciling the Fermi-LAT gamma-ray flux with neutrinos~\cite{Murase_39}. This tension could be resolved if the neutrino-producing sources were gamma-ray-opaque at GeV energies and above. The~diffuse neutrino flux can then be explained as coming from photohadronic interactions~\cite{Murase_46}.

Blazars have long been the focus of many neutrino searches, both time-dependent and stacked~\cite{blazar_stack1, blazar_stack2, sharma2021, Abbasi_2024}. Due to their powerful AGN engines and optimal conditions for particle acceleration, coupled with their strong beaming, their quiescent and active (flaring) states are both assumed to be capable of contributing to the high-energy diffuse flux of neutrinos~\cite{blazar_flare1, blazar_flare2, kun, Kun_2023}. Due to a dense environment of target photons, \textit{$p\gamma$} is expected to be the relevant channel of neutrino production in blazars. Whether they contribute to the PeV range or above is still an open question, and~one tied to the production mechanisms of UHECRs~\cite{murase_50, murase_108}. 

Direct experimental evidence, however, to~support the case for blazars was scarce prior to 2017. A~multi-messenger follow-up campaign, post an IceCube track-like alert, led to the first strong spatial and temporal correlation between a blazar and a high-energy neutrino~\cite{txs_ic}. TXS 0506+056, at~a redshift of  z = 0.33, was in the middle of a months-long gamma-ray flare when the neutrinos ($E_{\nu} \sim 290$ TeV) arrived from its direction. Fermi-LAT and MAGIC~\cite{ansoldi_2018} observed the GeV excess, while Swift and NuSTAR confirmed the X-ray activity~\cite{txs_ic}. IceCube performed an archival data fit in the source direction, which found an excess below 100 TeV with~an independent significance of $3.5 \sigma$~\cite{txs_nuflare}. 

TXS 0506+056 is an intermediate-peaked blazar, at~a moderate redshift, with~a bolometric luminosity $L_{bol}$$\sim$1.7 $\times 10^{45}$ erg/s. It was initially classified as a BL Lac, but recent investigations based on its luminosity and optical line properties have changed its status to an intrinsic FSRQ masquerading as a BL Lac~\cite{txs_masq}. The~almost-simultaneous observations in multiple wavelength bands during the $\gamma$-ray flare allowed for a comprehensive multi-messenger modeling of the SED during that period and~tightly constrained the neutrino flux and emission scenarios. One-zone lepto-hadronic models fail to reconcile the X-ray data with the neutrino expectations for the 2017 flare~\cite{Keivani_2018}. The~archival neutrino flare did not have an X-ray or gamma-ray counterpart, and~the single-zone models in agreement with the neutrino excess within the period overshoot the observations of where the cascaded flux from hadronic gamma-rays is expected to lie~\cite{Murase_115, Murase_116, Murase_117, Murase_118}. Models with neutral beams and multiple emission zones have been proposed that can account for the EM and neutrino data~\cite{Murase_115, Murase_119, Murase_120}; however, no single favored scenario has emerged.  

\begin{figure}[H]
     
     \begin{subfigure}[b]{0.45\textwidth}
         
         \includegraphics[width=\textwidth]{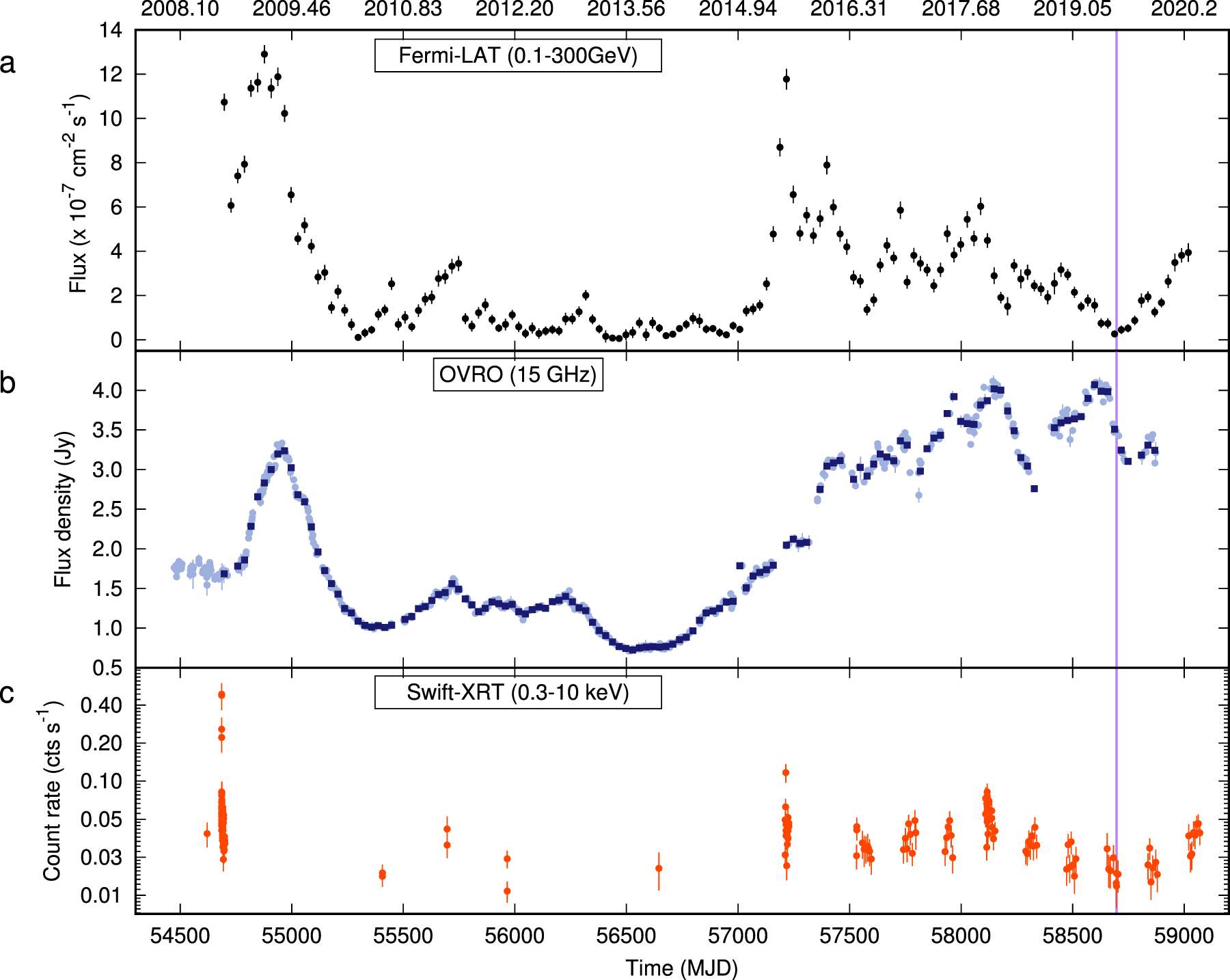}
         \label{fig:y equals x}
     \end{subfigure}
     \hfill
     \begin{subfigure}[b]{0.52\textwidth}
         
         \includegraphics[width=\textwidth]{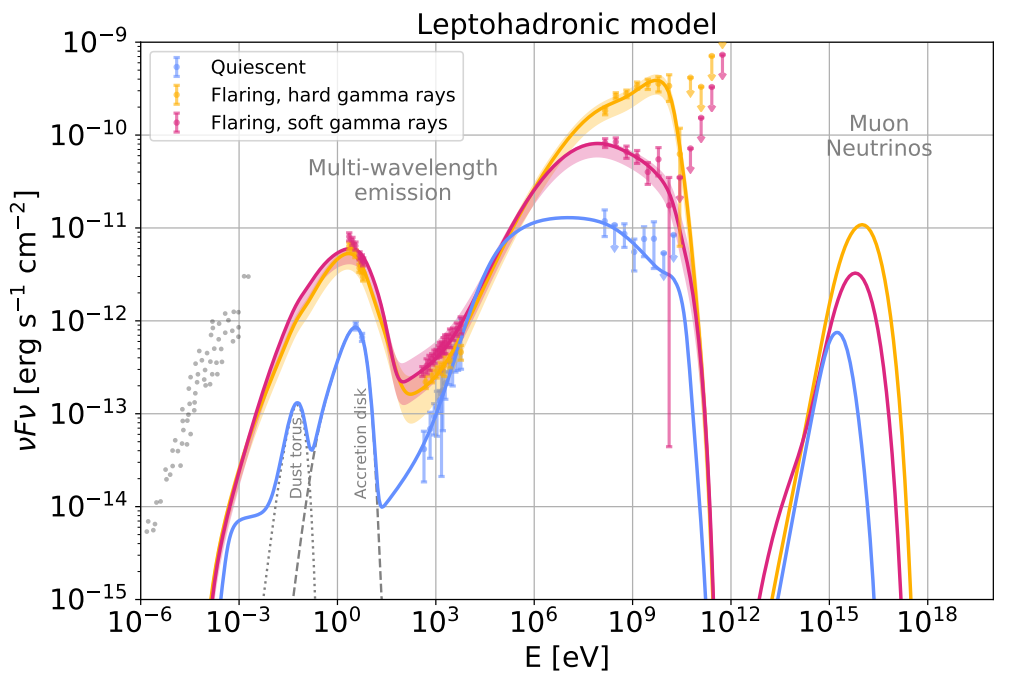}
         \label{fig:five over x}
     \end{subfigure}
        \caption{The multi$-$messenger behavior of FSRQ PKS 1502+106. Left panel: Fermi-LAT, OVRO and SWIFT light curves of the source, from~\cite{kun}. {The vertical line corresponds to the time of detection of the IceCube neutrino in coincidence with the FSRQ.} Anti-correlation is evident between the radio and $\gamma$-ray data. Right panel: multi-messenger modeling of the source, using a leptohadronic model by~\cite{Rodrigues_2021}, during~its quiescent and flaring states. The~model fit indicates that the radio emission may originate from a different area than the neutrinos and X-rays, and~hence be~uncorrelated. }
        \label{fig:pks}
\end{figure}

The perplexing case of the lone neutrino flare from TXS 0506+056, and~the inevitable connection between its neutrino flux and cascaded photon flux, has inspired the investigation of neutrino associations with blazars showing temporarily suppressed $\gamma$-ray emission and increased or correlated radio and/or X-ray activity~\cite{kun}. Multi-messenger modeling during the flaring of one of these blazars, PKS 1502+106, an~FSRQ at a redshift of $z = 1.84$, strongly suggests a hadronic origin for the soft X-ray flux from the source~\cite{Rodrigues_2021}. Moreover, OVRO~\cite{ovro} data on the source shows an anti-correlation trend between radio and $\gamma$-rays around the time of neutrino coincidence, with~a flare in the radio band coinciding with the $\gamma$-ray suppression~\cite{kun}, as~seen in Figure~\ref{fig:pks}.

Separately from hard and soft X-rays and $\gamma$-rays, radio-bright AGN have also been a focus of multi-messenger correlation searches~\cite{ic_radio_nu, ic_radio_nu1, desai2021testing}. Disentangling the region of emission within a source is not possible solely with neutrinos due to the limited angular resolution of the current, and even proposed, neutrino observatories. However, a~strong correlation between neutrino directions and radio-bright sources may suggest that neutrino acceleration sites are more likely located close to the parsec-scale core of the AGN (the accretion disks, jet launching regions, etc.), rather than extended blobs and hotspots farther out in the galaxy. 

Another AGN that made recent headlines for being the second extra-galactic source of high-energy neutrinos was the Seyfert galaxy NGC 1068~\cite{ic_ngc}. Although~the $4.2 \sigma$ evidence accumulated in its favor came from looking at IceCube data alone, EM observations have helped narrow down the possible emission scenarios. Studies have already indicated that the AGN at its center, and~not the starburst activity within the galaxy, might be responsible for neutrino production~\cite{padovani2024supermassiveblackholeshighenergy, murase_ngc_agn}. 

Hints for AGN--neutrino associations have continued to accumulate, with most of them being blazars in spatial coincidence with the IceCube alerts and showing simultaneous flaring activity in other bands~\cite{Murase_121, Murase_122, Murase_123, Rodrigues_2021, alerts_corr3}. But,~as of present, there is not sufficient observational evidence to conclusively quantify their contribution. It is also unclear whether this contribution comes from a few gamma-ray-bright AGN or many faint sources, or~if these sources contribute sparingly over time or emit a constant flux on average. 

Additionally, almost all significant LIGO/{Virgo} detections~\cite{gw_first, ligo_2017} have been followed up by neutrino observatories {\cite{ligo_ic1, ligo_ic2, ligo_antares, ligo_km3net1, Albert_2017}}, but~none have returned a positive association to date, despite the much higher success achieved in identifying their EM counterparts. The~cannonball model of GRBs can be evoked to explain the missing counterparts, wherein the merger results in two collimated jets that are beamed very narrowly, resulting in only few of them possibly lining up with Earth~\cite{nu_gw_flux}. This lack of detection has set strong limits on these transients, demonstrating that, away from the jet line of sight, binary mergers are not expected to be very high neutrino fluence events~\cite{ligo_ic}. 

Correlating the arrival directions of high-energy neutrinos with UHECRs has also failed to show any significant excess. Recent studies with larger datasets have reduced the level of overfluctuations observed previously~\cite{nu_cr}. The~extremely low flux of UHECRs makes it challenging to collect enough statistics for a thorough investigation of their origin. However, the CRs of energies a few orders of magnitude smaller can also be expected from inside our own galaxy, and~with a slightly higher flux. With~the new neutrino map of the galactic plane released by IceCube~\cite{galactic_ic}, it might be possible to identify galactic PeVatrons and improve our models of CR propagation within the galaxy in the near future through multi-wavelength and multi-messenger connection.

\section{Future~Outlook}

Efforts made in the field of neutrino astronomy over the last couple of decades have provided several tantalizing hints about where to and how to look for astrophysical neutrino sources. Put together, these milestones have paved the way for future missions that make improvements in sensitivity or fill in the gaps to increase our observational range. Multi-messenger collaboration is one such strategy. It benefits from advancements in EM and neutrino observation capabilities and a streamlined flow of information between the two~sectors.

\begin{figure}[H] 

\includegraphics[width=10cm]{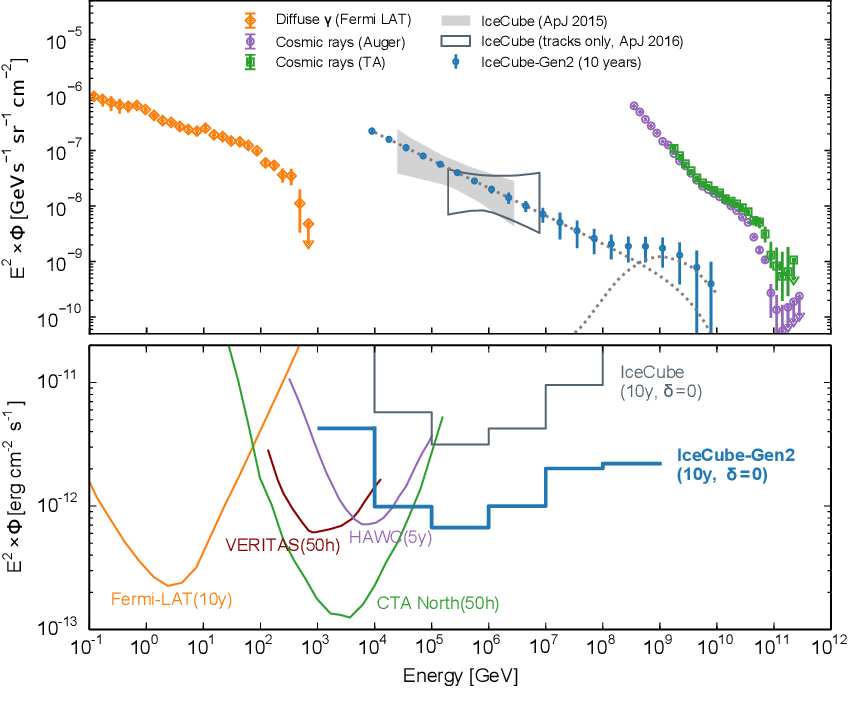}
\caption{IceCube$-$Gen2's performance over 10 years, seen under a multi-messenger lens. Upper panel: comparison of Fermi-LAT's diffuse $\gamma$-ray flux~\cite{ahlers100} with the neutrino flux from  IceCube tracks~\cite{gen2_84, gen2_85} and expectations from IceCube-Gen2 (assuming a continuing spectrum of $\Phi \propto$ $E^{-2.5}$ and~a cosmogenic neutrino flux with a $10\%$ proton fraction in UHECRs~\cite{gen2_53}, indicated by gray dotted lines), as well as the cosmic ray flux at ultra-high energies from Auger~\cite{gen2_83} and TA~\cite{gen2_82}. Lower panel: IceCube and IceCube-Gen2's (optical + radio array) differential sensitivity in 10 years, at the declination of the celestial equator, compared to that of Fermi-LAT, VERITAS, HAWC and CTA North. For~details on calculations, refer to~\cite{ic_gen2_mm}. Panel adapted from~\cite{ic_gen2_mm}.}
\label{fig:gen2}
\end{figure}

The first and foremost of these efforts lies in improving our capability to detect neutrinos. This can be achieved in a few ways: by increasing the instrumented area for detection, reducing background contamination and~expanding sky coverage. IceCube-Gen2 is proposed to cover $\sim$8 km$^3$ and increase the energy flux sensitivity (between \mbox{0.1 and 1 PeV}) by a factor of $\sim$10, bringing it to the level of $10^{-12}$ erg/cm$^2$/s within a few years of operation~\cite{ic_gen2}. Notably, this is comparable to the current flux limits of the VHE gamma-ray observatories like HAWC,  VERITAS and CTA (Figure \ref{fig:gen2}). KM3NeT/ARCA, GVD-Baikal and P-ONE will complete their construction within this decade and, together with IceCube, will form a GNN. Each of these GigaTon-volume detectors will cover complementary regions of the sky and provide alerts to the community without blind spots. Additionally, possible coincident observations across different observatories will strengthen the significance and reduce the search window for EM follow-ups.

Another important factor that constrains follow-up observations is the uncertainty region in the sky for neutrino events. While a sub-degree angular error for IceCube alerts is common, source confusion still persists. This can be addressed by better event reconstruction and more sensitive instrumentation. Machine learning plays, and~will continue to play, an important part in enhancing our reconstruction algorithms. {Graph Neural Networks and Convolutional Neural Networks both have been explored in the context of improving the reconstruction of cascade and track events~\cite{cnn_ic, cnn_antares, gnn_km3, gnn_ic} compared to classical approaches, while Graph Neutral Networks are also being favored for signal classification in IceCube~\cite{gnn_ic_clf}.} Understanding the detector medium's characteristics, as~well as the systematic uncertainty involved in reconstruction~\cite{ic_systematics}, is another important criterion for refining event reconstruction in neutrino telescopes and will consequently make multi-messenger campaigns more effective. Multi-PMT Digital Optical Modules~\cite{m-DOM}, like those in the KM3NeT detector, can constrain the optical background to a greater degree, improving sensitivity in the sub-TeV range, while at the same time the directional discrimination of collected light provides a better angular resolution at higher energies~\cite{km3net_loi}.

\begin{figure}[H] 

\includegraphics[width=8cm]{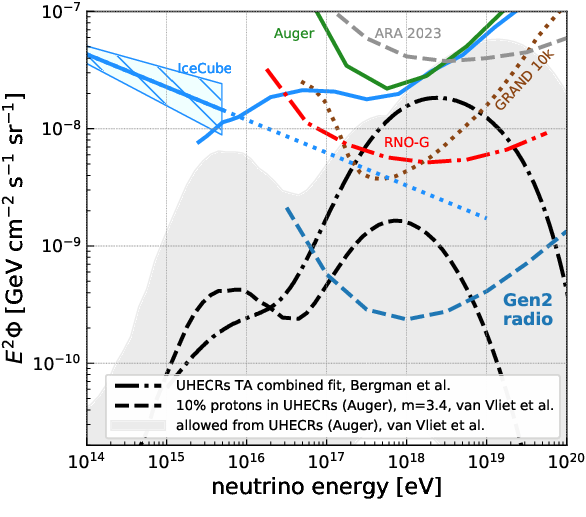}
\caption{The astrophysical neutrino flux~\cite{gen2_radio_1} observed by IceCube, extended to higher energies and plotted against the 10 year 90\% C.L. differential sensitivity of the Gen2-radio array. The expected sensitivities of ARA (2023), RNO-G and GRAND 10k (10 years) are also shown. {Black segmented lines are predictions of the cosmogenic neutrino flux based on TA and Auger measurements of UHECRs' spectrum and composition.} Figure from~\cite{gen2_radio_2}, with~permission.}
\label{fig:radio_sensi}
\end{figure}

\begin{figure}[H] 

\includegraphics[width=12cm]{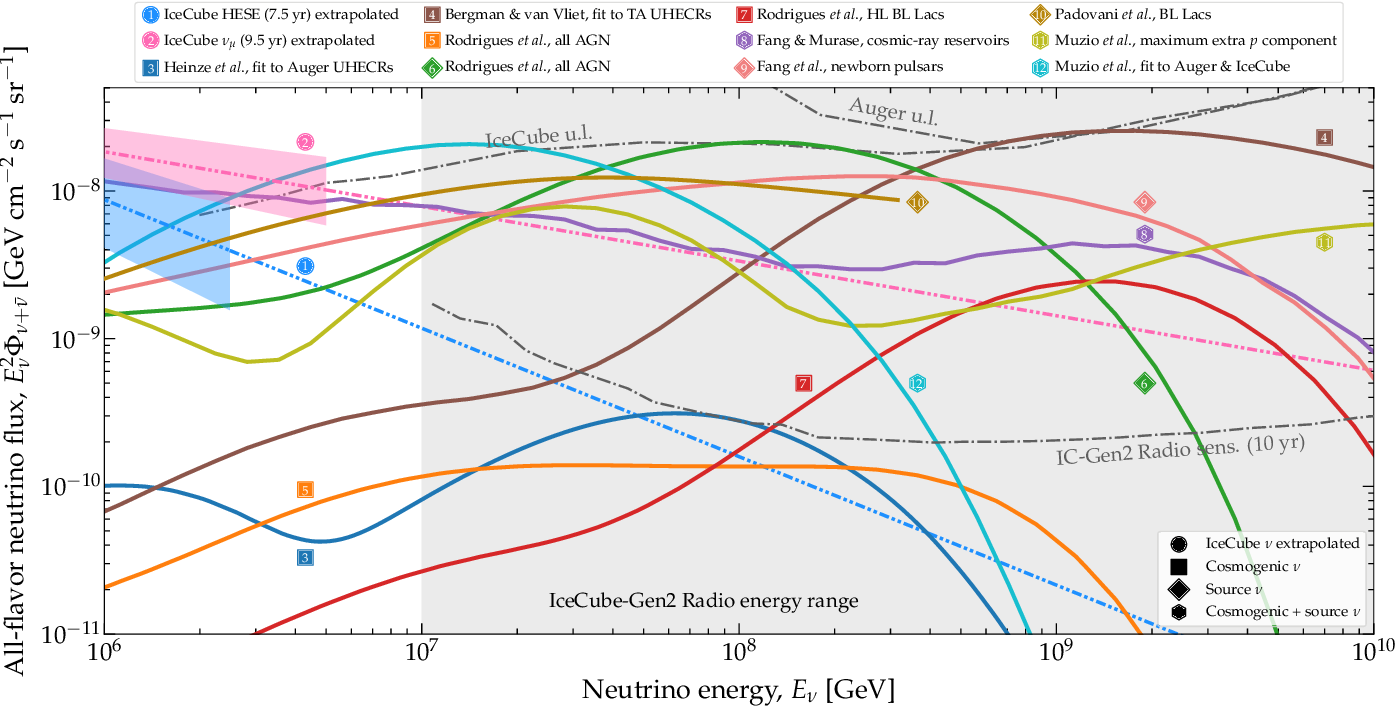}
\caption{Sensitivity of IceCube$-$Gen2 radio array~\cite{ic_gen2} compared to the diffuse neutrino flux models in the literature. The upper limits of IceCube~\cite{murase_50} and Pierre-Auger~\cite{auger_gen2} are also shown. For~model details, see the original paper~\cite{glaser}. }
\label{fig:agn_gen2}
\end{figure}

{IceCube-Gen2 also comprises a radio component which is separate from its optical extension, and uses the Askaryan effect for the radio detection of high-energy neutrinos~\cite{askaryan}. This radio array will vastly increase the observable energy range}, bringing the EeV neutrino sky, and~hence cosmogenic neutrino flux, within~reach (Figure \ref{fig:radio_sensi}) \cite{gen2_radio}. Combined with its extended surface array for vetoing air showers, IceCube itself will be a multi-messenger observatory. {It will be complemented by other EeV-sensitive radio neutrino observatories like RNO-G in the far north~\cite{rno} and the mid-latitude GRAND array~\cite{grand}, along with the POEMMA space mission~\cite{poemma}.} This significant increase in reach over decades of energy will make it possible to address the question of which classes of AGN contribute more below the PeV range and which contribute to the PeV--EeV flux of the neutrino spectrum \mbox{(\mbox{Figure \ref{fig:agn_gen2}}) \cite{blazar_flare2, flux_model1, flux_model2, flux_model3}}. {Alert programs are planned for each of these observatories and,~with the sub-degree angular resolution envisaged for GRAND and POEMMA, neutrino-transient follow-ups will also be possible at EeV energies~\cite{foteni_review}.}

\unskip

High-energy-neutrino-emitting sources are expected to emit over a wide range of the EM spectrum, necessitating the need for coverage over a wide band of wavelengths. Neutrino telescopes are high duty cycle detectors, optimally covering almost half the sky at any given time. As~such, correlation studies in the EM regime also require wide-field-of-view instruments and high-cadence surveys of the sky across the energy range in question, which extends from radio to TeV. Adding source variability to this equation dictates also the requirement for the monitoring of candidate sources, for~a reference on the flaring states vs. the quiescent periods of these sources. The~{MOJAVE} program~\cite{mojave} is one such effort in the radio sector. Putting all the pieces into place will increase the frequency of ToO observations and their rate of success.

Space and ground-based missions are planned to address the current blind spots in our EM sky. OVRO and MeerKAT~\cite{meerkat} have been consistently providing radio data on AGNs over the past decade. SKA~\cite{ska} will finish construction in 2028 and aid in refined observations of galactic sources. {Additionally, it will also map the IGMF and galactic magnetic field in great detail, along with that of individual galaxies like AGN and starbursts~\cite{ska_magnetism}. This will shed light on the acceleration sites within these sources and the processes involved, thus improving the characterization of potential sources of cosmic ray acceleration. ngEHT~\cite{ngEHT} and ngVLA~\cite{ngVLA} will be next-generation additions that will play a complementary role in this regard. ngVLA will search for the radio counterparts of neutrino or GW transients and ngEHT will be capable of mapping the accretion disk magnetic field and resolving the cores of AGN.} The microwave range will benefit from the ground-based CMB-S4 experiment~\cite{cmb_s4}, which will make it possible to perform transient follow-ups in the $\upmu$m band as well. 

Optical counterparts to transients such as GRBs and TDEs are currently studied using facilities like the ZTF, PanSTARRS~\cite{panstarrs}, etc. This already vast landscape will be revolutionized by the Vera Rubin Observatory, with~an unprecedented number of nightly alerts for transients. New transient classes with potential significance for high-energy neutrino associations might also possibly emerge from this zoo of alerts. Wide optical surveys like the LSST by Vera Rubin, and~those by Euclid~\cite{euclid} and the Roman Space Telescope~\cite{roman}, will be highly advantageous for categorizing the source class of neutrino emitters, especially star-forming galaxies. 

X-ray and gamma-ray data become critical for modeling neutrino emissions from the point sources of neutrinos, since the high-energy peak is where the clues for disentangling hadronic emission are hidden in these vigorous non-thermal emitters. The~cascades from hadronic gamma-rays are expected to end up in the keV--MeV band. The~{SWIFT} observatory~\cite{swift}, which succeeded the ROSAT experiment, has provided valuable insights on transients in the soft X-ray ($<$10 keV) band with its fast pointing capability. MAXI and INTEGRAL also cover this band and~have wider fields of view. However, e-ROSITA's\endnote{This satellite stopped taking data at the start of the Ukraine--Russia war in February 2022, although~its first all-sky survey data was recently released.} all-sky surveys~\cite{erosita_old, Merloni_2024} could have been hugely beneficial for X-ray investigations. With~its large field of view and high-cadence monitoring,  it would have allowed us to produce long-term X-ray light curves akin to Fermi-LAT. NuSTAR remains the most sensitive instrument in the hard X-ray band, while the MeV band remains a blind spot as of now. This should, however, soon be fixed by upcoming missions in various stages of implementation, like  AMEGO-X~\cite{Caputo_2022} and e-ASTROGAM~\cite{De_Angelis_2018}. {These two space missions will roughly cover the 100 keV--3 GeV band, the~energy scales where the high-energy bump of most AGN is expected to fall. AMEGO-X will view the entire sky every two orbits to~compile a detailed map of the MeV--GeV gamma-ray sky, while e-ASTROGAM's wide FoV will allow for the monitoring of variable sources. e-ASTROGAM will also be able to provide polarimetric data for these sources and,~coupled with its high angular resolution (0.15$^{\circ}$ at 1 GeV), will help mitigate source confusion in follow-ups.} 

The Fermi-LAT is sensitive in the 100 MeV--500 GeV band and~monitors the entire gamma-ray sky every 3 h with its wide field of view. It is complemented by the GBM, a~fast-pointing instrument sensitive in the  keV-MeV band for following up transients in real time and issuing alerts for the same. Fermi-LAT overlaps in coverage with the VHE $\gamma$-ray observatories like HESS, MAGIC and VERITAS in the lower GeV range.
In~the next few years, CTA~\cite{CTA} will supersede HESS and,~with its much larger collection area, will be able to detect neutrino counterparts  much farther away than the current facilities. {It will also boast of an arc-minute angular resolution at higher energies, crucial for resolving emission sites in extended sources and for identifying counterparts in source-dense regions of the sky.} Along with LHAASO~\cite{lhaaso} and HAWC~\cite{HAWC:2023amj}, and~the upcoming SWGO~\cite{Chiavassa:2024rew} in the Southern Hemisphere, our monitoring capabilities will also be significantly improved. {These VHE observatories will together improve the multi-messenger modeling paradigm by being more capable of observing spectral cutoffs in SEDs. LHAASO in particular will be extremely critical for the UHE follow-ups of galactic transients. It has already, in its current configuration, observed Galactic PeVatrons~\cite{lhaaso_catalog} and an SNR that hints at pion decay features in its spectrum~\cite{lhaaso_snr}.}  

High-power transient astrophysical phenomena like GRBs, PWN, TDEs, BH mergers, blazar flares, etc., have been comprehensively studied to predict their high-energy neutrino expectations~\cite{trans1, trans2, trans3, trans4, trans5, trans6, trans7, trans8, Abbasi_2024, trans10}. Multi-messenger campaigns at all wavelengths are key to identifying the neutrino signals from transients~\cite{foteni_review}. At~the same time, the time-scales and frequency of these transients will provide the testing grounds for the next-generation multi-messenger networks and alert brokers like AMON, SciMMA, TACH, AMPEL~\cite{ampel}, SNEWS 2.0, etc., discussed in Section~\ref{sec:realtime}. The~existing MoU-based multi-messenger networks like the ones for OFU and GFU alert dispersion will continue to be relevant and~will proliferate, with the inclusion of the next generation of observatories in their respective wavelength regimes. 

Neutrino astronomy is set to peak in the coming decades. A~significant contribution to its success in changing the neutrino sky paradigm will lie in its ability to co-evolve with other dedicated observatories around the world for every messenger and specific wavelength bands, as~they fill in the gaps in our observational capabilities. Technological advances in the tools and instruments in use, and~the continued integration of multi-messenger data, will refine our models and proliferate breakthroughs and discoveries, offering a more cohesive and comprehensive view of the high-energy universe.

\vspace{6pt} 





\funding{This research was supported by the Data Intelligence Institute of Paris (diiP) and IdEx Université Paris Cité.}



\acknowledgments{{The author would like to thank Yvonne Becherini (APC, Paris) for proofreading the article and her insightful comments. Many thanks also to Camilla Gönczi for her assistance with designing and improving the figures.}}

\conflictsofinterest{The author has no conflicts of interest to~declare.} 



\abbreviations{Abbreviations}{
The following abbreviations are used in this manuscript (ordered alphabetically):\\

\noindent
\begin{tabular}{@{}ll}
AGN & Active Galactic~Nucleus \\
AI & Artificial~Intelligence \\
AMEGO-X & All-sky Medium-Energy Gamma-ray Observatory~eXplorer \\
AMON & Astrophysical Multi-messenger Observatory~Network \\
AMPEL & Alert Management, Photometry, and~Evaluation of Light~curves \\
ANTARES & Astronomy with a Neutrino Telescope and Abyss Environmental~RESearch \\
ARA & Askaryan Radio~Array \\
ARCA & Astroparticle Research with Cosmics in the~Abyss \\
ASAS-SN & All-Sky Automated Survey for~SuperNovae \\
Baikal-GVD & Baikal Gigaton Volume~Detector \\
BBH & Binary Black~Hole \\
BH & Black~Hole\\
BL Lac & BL Lacertae~Object \\
CMB-S4 & Cosmic Microwave Background Stage~4 \\
CRs & Cosmic~Rays \\
CTA & Cherenkov Telescope~Array \\
DOAJ & Directory of open access~journals\\
e-ASTROGAM & Enhanced~ATROGAM \\
EBL & Extragalactic Background~Light \\
EeVs & Etta electron volts ($10^{18}$ electron volts) \\
EHE & Extremely High-energy~Events \\
EM & Electromagnetic\\
eROSITA & Extended ROentgen Survey with an Imaging Telescope~Array \\
Fermi-LAT & Fermi Large-Area~Telescope \\
FoV & Field of~View \\
\end{tabular} 

\noindent
\begin{tabular}{@{}ll}
FRA & Fast Response~Analysis \\
FSRQ & Flat Spectrum Radio~Quasar \\
GCN & General (Gamma-Ray) Coordinates~Network \\
GeV & Giga electron volts ($10^9$ electron volts) \\
GFU & Gamma-ray Follow-Up \\
GNN & Global Neutrino~Network \\
GRAND & Giant Radio Array for Neutrino~Detection \\
GRB & Gamma-Ray~Burst \\
GW & Gravitational~Wave \\
HAWC & High-Altitude Water Cherenkov~Observatory \\
HESE & High-Energy Starting~Events \\
HESS & High-Energy Stereoscopic~System \\
IGMF & Inter Galactic Magnetic~Field \\
INTEGRAL & INTErnational Gamma-Ray Astrophysics~Laboratory \\
KM3NeT & KiloMeter$^3$ Neutrino~Telescope \\
LHAASO & Large High-Altitude Air Shower~Observatory \\
LIGO & Laser Interferometer Gravitational-wave~Observatory \\
LLM & Large Language~Models \\
LOFAR & LOw Frequency~Array \\
LSST & Large Synoptic Survey~Telescope \\
MAGIC & Major Atmospheric Gamma Imaging Cherenkov~Telescopes \\
MASTER & Mobile Astronomical System of Telescope Robots \\
MAXI & Monitor of All-sky X-ray~Image \\
MDPI & Multidisciplinary Digital Publishing~Institute\\
MeerKAT & Karoo Array~Telescope \\
MeV & Mega electron volts ($10^{6}$ electron volts) \\
MOJAVE & Monitoring Of Jets in Active galactic nuclei with VLBA~Experiments  \\
NGC & New General~Catalogue \\
ngEHT & Next-Generation Event Horizon~Telescope \\
ngVLA & Next-Generation Very Large~Array \\
NuSTAR & The Nuclear Spectroscopic Telescope~Array \\
OFU & Optical and X-ray Follow-Up \\
OVRO & Owens Valley Radio~Observatory \\
PanSTARRS & Panoramic Survey Telescope \& Rapid Response~System \\
PeV & Peta electron volts ($10^{15}$ electron volts) \\
PMTs & Photo-Multiplier~Tubes \\
POEMMA & Probe of Extreme Multi-Messenger~Astrophysics \\
P-ONE & Pacific Ocean Neutrino~Experiment \\
PWN & Pulsar Wind~Nebulae \\
RNO-G & Radio Neutrino Observatory in~Greenland \\
ROSAT & Röntgen~Satellite \\
SciMMA & Scalable Cyberinfrastructure to Support Multi-Messenger~Astrophysics \\
SED & Spectral Energy~Distribution \\
SKA & Square Kilometer~Array \\
SN & SuperNova \\
SNEWS & SuperNova Early Warning~System \\
SNRs & SuperNova~Remnants \\
SWGO & Southern Wide-field Gamma-ray~Observatory \\
TA & Telescope~Array \\
TACH & Time-Domain Astronomy Coordination~Hub \\
TDE & Tidal Disruption~Event \\
TeV & Tera electron volts ($10^{12}$ electron volts) \\
ToO & Target of~Opportunity \\
UHE & Ultra-High~Energy \\
UHECRs & Ultra-High-Energy Cosmic~Rays \\
VERITAS & The Very Energetic Radiation Imaging Telescope Array~System \\
VHE & Very High~Energy \\ 
VLA & Very Large~Array \\
XMM-Newton~~& X-ray Multi-mirror Mission-Newton \\
ZTF & Zwicky Transient~Facility 

\end{tabular} 
}

\appendixtitles{no} 




\begin{adjustwidth}{-\extralength}{0cm}
\setenotez{list-name=Note}
\printendnotes[custom] 

\reftitle{References}

\PublishersNote{}
\end{adjustwidth}
\end{document}